\documentclass[a4paper,12pt]{article}
\usepackage{hyperref}
\usepackage{tikz,pgfplots,pgfplotstable,filecontents}
\pgfplotsset{compat=1.17}
\usetikzlibrary{arrows,positioning,calc,fadings,decorations.pathreplacing,decorations}
\usepgfplotslibrary{patchplots,colormaps}
\usepackage{amssymb,amsmath,amsthm,mathtools}
\usepackage{subfig,caption}
\usepackage{cite}
\usepackage{enumerate}
\usepackage[affil-it]{authblk}
\newcommand\sChannel{
\begin{tikzpicture}[x=4pt,y=2pt]
	\draw (-.5,1) -- (0,0);
	\draw (-.5,-1) -- (0,0) -- (1,0) -- (1.5,1);
	\draw (1,0) -- (1.5,-1);
\end{tikzpicture}}
\newcommand\tChannel{
\begin{tikzpicture}[x=2pt,y=4pt]
	\draw (-1,1) -- (0,0.5);
	\draw (1,1) -- (0,0.5) -- (0,-.5) -- (-1,-1);
	\draw (1,-1) -- (0,-.5);
\end{tikzpicture}}

\def\R{\mathbf{R}}

\def\mA{\mathcal{A}}

\newcommand\Ga[1]{\Gamma\left(#1\right)}

\newcommand\eq[1]{(\ref{#1})}
\def\pa{\partial}

\def\m{{\boldsymbol\mu}}
\def\n{{\boldsymbol\nu}}
\def\bxi{{\boldsymbol\xi}}

\newcommand\rt{\longrightarrow}
\def\I{\mathcal{I}}

\theoremstyle{definition}

\def\m{\boldsymbol{\mu}}
\makeatletter
\newcommand\inmu[3]{I^{{#1}}_{{#2}}({#3})}
\makeatother

\def\D{\Delta}
\def\bD{\boldsymbol{\Delta}}
\def\bxi{\boldsymbol{\xi}}

\newcommand\phiD[2]{\phi_{#1}( {#2})}
\newcommand\corr[3]{G^{(#1)}_{#2}(#3)}
\def\x{\boldsymbol{x}}

\newcommand\braket[1]{\left\langle#1\right\rangle}

\title{Conformal bootstrap in Mellin space from GG systems}
\author{Koushik Ray\thanks{koushik@iacs.res.in}}
\affil{Indian Association for the Cultivation of Science,\\
Kolkata - 700032, India.}
\begin{document}
%
%
%
%
%
\maketitle
\thispagestyle{empty}
\begin{abstract}
\noindent
A simple scheme to express the Mellin transform of 
$D$-dimensional Euclidean conformal bootstrap equation is presented by 
relating conformal blocks to
a Gauss-Grassmann (GG) system due to Gelfand-Graev,
associated to conformal integrals, which, in turn, are generalised
hypergeometric functions.
Usefulness of the expression for obtaining bounds
on the spectrum of fields is demonstrated.
\end{abstract}
Specification of a conformal field theory (CFT) entails 
providing the spectrum of primary fields in terms of
quantum numbers associated to the generators of the conformal algebra and 
the constants in the ring of fields, called the operator product expansion 
(OPE) coefficients. Correlation functions of fields of 
a CFT depend on these data. 
The conformal bootstrap program is a scheme which seeks to restrict these data
\cite{FGG,Pol,mack,bpz}.
Inequivalent permutations of a configuration of fields are called channels. 
Obtaining restrictions on OPE coefficients 
as well as the spectrum of fields by equating  correlation functions 
of configurations of $N$ points, called $N$-point functions, in 
different channels is the desideratum. 
The bootstrap program has been extensively studied 
from its inception to the present day \cite{rychkov1}. 
Since the form of $2$- and $3$-point functions
are completely fixed by conformal covariance of the fields, such a scheme 
starts from the $4$-point functions. An $N$-point function for 
$N>3$ contains a piece, called the conformal block,
which is a function  of geometric conformal invariants of the spacetime
coordinates, called cross ratios. Since any function of the invariants is an 
invariant itself, these introduce a certain arbitrariness of the correlation
functions, which is fixed by demanding consistency with the representation
of the conformal group. 
While the conformal block is generally expressed as an infinite series as
(generalised) hypergeometric functions of cross ratios from the representation 
theory of the conformal group, certain details require going beyond 
\cite{monodrom}. Thus, the bootstrap equations 
are equations involving infinite series of cross ratios,
rendering them difficult to solve, to say the least. 
Furthermore,  the conformal weights, that is the
quantum numbers associated with the scaling generator of the conformal group, 
may be valued in a 
countable or an uncountable set of real or complex numbers. Correlation 
functions are expressed as a sum or integral 
over these weights, posing another hurdle
for the  bootstrap programme. 
A variety of techniques, both 
analytic and numerical, along with their combinations, have been devised
over the years to deal with these equations 
\cite{ratta,MP,CDKPS,CEHS,PRV,num1,dgs,aninda,functional,faiz}.
Most of these tackle the problem by integrating the bootstrap equation
over the cross ratios, perhaps with a weight function, by considering
a functional of conformal blocks. This is usually achieved by continuing the
real cross ratios to the domain of complex numbers which requires
settling extremely subtle issues of choosing integration contours in
conformity 
with Fubini's theorem \cite{QR}. A variant of this approach is to consider
Mellin transform of the bootstrap equation with respect to the cross ratios,
where the weight of integration is fixed, although the domain of 
convergence of the infinite series in the conformal block still poses 
subtleties \cite{rajesh,GKSS1,GKSS2}.  

In this article we set up the bootstrap equations for a conformal field
theory of scalar fields alone. We consider the Mellin transform of the 
equations with respect to cross ratios. Issues pertaining
to the convergence of infinite series in the conformal block is eschewed
by taking cognizance of the fact that the conformal block is written
in terms of a solution 
to the GG system of hypergeometric functions which has an integral 
presentation. The conformal correlation functions have been expressed
earlier in terms of conformal integrals which in turn are 
written as generalised GKZ hypergeometric functions and known in all
dimensions \cite{PR1,PR2,PR3}. 
While the conformal integral is covariant under conformal transformations,
it contains an invariant piece which is related to the conformal block.
Writing the conformal integral in terms of the solution to a GG system
allows expressing the conformal block 
as an integral, containing
the cross ratios as parameters \cite{gg1,gg2}. It is then easy to take the 
Mellin transform of it avoiding the difficulties associated
with the convergence of infinite series. 

In the rest of the article we explain the identification of  conformal 
blocks with solutions of GG systems, leading to the expression of
bootstrap equations in a straightforward manner. In order to keep
the discussion simple we demonstrate 
the idea for the $4$-point correlation function of a scalar conformal 
field theory. Generalisation to higher point functions is obviously
possible along the same line.

Let $\phiD{\D}{x}$ denote a $D$-dimensional primary scalar field 
with conformal weight $\D$, where  
$x\in\R^D$, the $D$-dimensional Euclidean space.
The Euclidean norm of $x$ is denoted $|x|$ and we write $x_{ij}=x_i-x_j$.
The $N$-point correlation function  is denoted as
\begin{equation}
\label{corr}
\corr{N}{\bD}{\x} = 
\braket{\phiD{\Delta_1}{x_1}\cdots\phiD{\Delta_N}{x_N}},
\end{equation}
where $\bD=(\D_1,\ldots,\D_N)$ is the $N$-tuple of weights of the fields and
$\x=(x_1,\ldots,x_N)\in(\R^D)^N$.
Covariance under conformal transformations fixes the $2$-point  
and $3$-point correlation functions up to numerical constants as
\begin{gather}
\label{G2}
\corr{2}{(\D,\D')}{x_1,x_2} = \frac{C_{\D}\delta_{\D\D'}}{|x_{12}|^{2\D}},
\\
\label{G3}
\corr{3}{(\D_1,\D_2,\D_3)}{x_1,x_2,x_3} = 
\frac{C_{\D_1\D_2\D_3}}{%
|x_{12}|^{\D_1+\D_2-\D_3}
|x_{13}|^{\D_1+\D_3-\D_2}
|x_{23}|^{\D_2+\D_3-\D_1}
},%
\end{gather}
where $C_\D$ is a constant, and $C_{\D_1\D_2\D_3}$ are the OPE coefficients. 
Defining the conformal integral
\begin{equation}
\label{inmu:def}
\inmu{\m}{N}{\x} = \int \frac{d^D x}{
|x-x_1|^{2\mu_1} |x-x_2|^{2\mu_2} \cdots|x-x_N|^{2\mu_N}},
\end{equation}
where $\m=(\mu_1,\mu_2,\ldots,\mu_N)$ is an $N$-tuple of weights  
with $|\m|:=\sum_{i=1}^N\mu_i=D$,  and redefining the constant
$C_\D$ appropriately,
the $4$-point correlation function 
$\corr{4}{(\D_1,\D_2,\D_3,\D_4)}{x_1,x_2,x_3,x_4}$ 
is written as \cite{PR1}
\begin{multline}
\label{G4s}
\corr{4\,\sChannel}{(\D_1,\D_2,\D_3,\D_4)}{x_1,x_2,x_3,x_4} =\\
\sum_{\D}
{C_{\D_1\D_2\D}C_{\D\D_3\D_4}}
\frac{\inmu{\left(\frac{\D_1-\D_2+\D}{2},
\frac{\D_2-\D_1+\D}{2},
\frac{\D_3-\D_4+D-\D}{2},
\frac{\D_4-\D_3+D-\D}{2}\right)
}{4}{x_1,x_2,x_3,x_4}
}{|x_{12}|^{\D_1+\D_2-\D}|\ x_{34}|^{\D_3+\D_4+\D-D}}
\end{multline}
in the $s$-channel.
In the $t$-channel the $4$-point correlation function 
assumes the form 
\begin{multline}
\label{G4t}
\corr{4\,\tChannel}{(\D_1,\D_2,\D_3,\D_4)}{x_1,x_2,x_3,x_4} =\\
\sum_{\D}
{C_{\D_1\D_4\D}C_{\D\D_3\D_2}}
\frac{\inmu{\left(
\frac{\D_1-\D_4+\D}{2},
\frac{\D_2-\D_3+D-\D}{2},
\frac{\D_3-\D_2+D-\D}{2},
\frac{\D_4-\D_1+\D}{2}
\right)
}{4}{x_1,x_2,x_3,x_4}
}{|x_{14}|^{\D_1+\D_4-\D}\ |x_{23}|^{\D_2+\D_3+\D-D}},
\end{multline}
obtained from \eq{G4s}
by simultaneous exchange  of $(x_2, \D_2)$ and $(x_4,\D_4)$.
The bootstrap equation is obtained by equating the correlation functions in
the two channels,
\begin{equation}
	\label{boot:def}
\corr{4\,\sChannel}{(\D_1,\D_2,\D_3,\D_4)}{x_1,x_2,x_3,x_4} =
\corr{4\,\tChannel}{(\D_1,\D_2,\D_3,\D_4)}{x_1,x_2,x_3,x_4},
\end{equation}
or, equivalently, presented diagrammatically, as,
\begin{equation}
	{\Huge\sum_{\D}} 
	\begin{tikzpicture}[x=4em,y=2em,baseline]
	\draw (-.5,1) node[above left] {$\scriptstyle \D_2$} 
-- (0,0) node[left]{$\scriptscriptstyle C_{\D_1\D_2\D}$};
		\draw (-.5,-1) node[below left] {$\scriptstyle \D_1$} -- 
		(0,0) to node[below] {$\D$} (1,0) node[right] 
{$\scriptscriptstyle  C_{\D\D_3\D_4}$}-- (1.5,1) node[above right] 
		{$\scriptstyle \D_3$};
		\draw (1,0) -- (1.5,-1) node[below right] 
		{$\scriptstyle \D_4$};
\end{tikzpicture}
=\quad
	{\Huge\sum_{\D}} 
\begin{tikzpicture}[x=2em,y=4em,baseline]
	\draw (-1,1) node[above left] {$\scriptstyle \D_2$} 
-- (0,0.5) node[right] {$\scriptscriptstyle C_{\D_2\D_3\D}$};
		\draw (1,1) node[above right] {$\scriptstyle \D_3$} -- 
		(0,0.5) to node[left] {$\D$} (0,-0.5) -- 
		(-1,-1) node[below left] 
		{$\scriptstyle \D_1$};
		\draw (0,-.5) node[right] {$\scriptscriptstyle C_{\D\D_1\D_4}$}
-- (1,-1) node[below right] 
		{$\scriptstyle \D_4$};
\end{tikzpicture}.
\end{equation}
In order to proceed further, the conformal integral is  expressed
as a generalised hypergeometric function 
which is a solution of a GKZ system. This allows writing the 
bootstrap equations in a particularly convenient form.
Let us briefly recall the association of hypergeometric functions to a
toric matrix. A toric matrix $\mA$ is a rectangular matrix
of positive integers of dimension, say, $m\times n$, $n>m$. 
Vectors annihilated by $\mA$ form an $(n-m)$-dimensional vector space. 
Independent elements of this space of null eigenvectors can be arranged into an
$n\times (n-m)$ matrix $G$, called the Gale matrix, so that $\mA {G}=0$.
Every column as well as any linear combination of the columns of the Gale 
matrix is annihilated by $\mA$. We denote the transpose of the Gale matrix 
by $L$. If the entries of $\mA$ are obtained from the exponents of scaling
of variables in  multivariate homogeneous polynomials, then each 
row of $L$ defines an element of an ideal in the set of such 
polynomials, thereby defining an algebraic variety. 
A non-zero number used for the scaling of variables 
is an element of an algebraic torus. Hence the name toric matrix.
The GKZ  system is a system of differential equations 
associated to the matrix $L$, which, in apropriate settings, yield the 
periods of the algebraic varieties as solutions. 
Associating the 
correlation functions to this geometric notion is beyond the scope of the
present article. 
The solutions of GKZ systems, on the other hand, can be looked upon as 
generalisation of the Gauss Hypergeometric function. Let us now set up 
the differential equations in our notations. 
The conformal integral transforms covariantly under the conformal 
transformations of $\R^D$. It is, therefore, a function of 
$|x_{ij}|$. It is a generalised hypergeometric function
satisfying the GKZ equations associated to the toric matrix
\begin{equation}
	\label{tormat:def}
	\mA_{i,(jk)} = \delta_{ij}+\delta_{ik},
\end{equation}
obtained from the weights of conformal transformations 
\cite{PR2,PR3}. The entry $\mA_{i,(jk)}$ in the toric matrix
is the exponent of scaling of $|x_{jk}|$ when $x_i$ is scaled
by a non-zero number, which is, of course, a conformal transformation. 
Cross ratios are scale invariants constructed as combinations of 
$|x_{ij}|^2$ 
with exponents given by the entries of a row of the matrix $L$. There are thus
as many cross ratios as the number of linearly 
independent null eigenvectors of $\mA$. Fixing a basis of the space of 
null eigenvectors
fixes a set of cross ratios, whose products are again scale invariants 
related to linear combinations of the basis vectors. 
In here, the toric matrix \eq{tormat:def}
is an $N\times N(N-1)/2$ matrix with $N(N-3)/2$ linearly independent
null eigenvectors, that is,
\begin{equation}
	\label{null}
\sum_{\substack{j,k=1\\j<k}}^N\mA_{i,(jk)}\ell^A_{jk} =0,
	\quad A=1,2,\ldots,N(N-3)/2.
\end{equation}
The $\ell^A$ form the rows of the matrix $L$.
The GKZ equations associated to this data are
\begin{gather}
	\label{gkz1}
	\left(	\prod\limits_{\ell^A_{ij}>0} \pa_{ij}^{\ell^A_{ij}}
- \prod\limits_{\ell^A_{ij}<0} \pa_{ij}^{\ell^A_{ij}}\right)\inmu{\m}{N}{\x}
	=0,\\
	\label{gkz2}
	\sum\limits_{j,k=1}^N \mA_{i,(jk)} |x_{jk}|^2\pa_{jk} \inmu{\m}{N}{\x}
	+\mu_i \inmu{\m}{N}{\x} =0,\quad\forall i=1,2,\ldots,N,
\end{gather}
where we defined $\pa_{ij}=\frac{\pa}{\pa |x_{ij}|^2}$.
As a solution of the GKZ equations the 
conformal integral \eq{inmu:def} is written as \cite{PR3}
\begin{equation}
	\label{IbI0}
	\inmu{\m}{N}{\x} = \prod\limits_{\substack{i,j=1\\i<j}}^N 
	|x_{ij}|^{2\beta_{ij}} \inmu{0}{N}{\bxi},
\end{equation}
where $\bxi=\{\xi^A; A=1,2,\ldots,N(N-3)/2\}$ are real-valued 
conformal invariant cross ratios, with 
\begin{equation}
	\label{xi:def}
	\xi^A = \prod\limits_{\substack{i,j=1\\i<j}}^N |x_{ij}|^{2\ell^A_{ij}},
\end{equation}
and $\beta$'s are parameters related to $\m$,  satisfying  
\begin{equation}
	\begin{gathered}
		\beta_{ii}=0,\ 
		\beta_{ji} = \beta_{ij},\\
		\label{betasum}
	\sum\limits_{\substack{i,j=1\\i<j}}^N 
		\mA_{i,(jk)}\beta_{jk} = -\mu_i,
	\end{gathered}
\end{equation}
$i,j,k=1,2,\ldots,N$.
By virtue of \eq{null}, the conformal integral \eq{IbI0} remains
unaltered under a shift of $\beta$'s by arbitrary multiples of the null 
vectors, \emph{viz.}
\begin{equation}
	\label{betashift}
	\beta_{ij} \rt \beta_{ij} + \sum_A n_A \ell^A_{ij}.
\end{equation}
The invariant part $\inmu{0}{N}{\bxi}$ of the conformal integral 
solves the GG system and is given by \cite{gg1}
\begin{equation}
	\label{I0GG}
	\inmu{0}{N}{\bxi} = \int\limits_{0\leqslant t_{ij}<\infty} 
\prod_{\substack{i,j=1\\i<j}}^N
	\frac{dt_{ij}}{t_{ij}} t^{-\beta_{ij}}_{ij} e^{-t_{ij}}
	\prod_{A=1}^{N(N-3)/2} 
	\delta\left(
	\frac{\prod\limits_{\substack{i,j=1\\i<j}}^N
	{t_{ij}^{\ell^A_{ij}}}}{\xi^A}-1
	\right).
\end{equation}
Although there are $N(N-1)/2$ number of $\beta$'s satisfying $N$ equations 
\eq{betasum},
the conformal integral \eq{inmu:def} written as \eq{IbI0} is independent
of the choice of $\beta$'s \cite{gg1,PR3}.
By the same token, 
the invariant $\inmu{0}{N}{\bxi}$ depends on the choice of $\beta$'s. 
The independence on the $\beta$'s is important, since the 
specification of a CFT and, therefore the conformal integral, makes no
allusion to them. The conformal integral contains the parameters $\m$
which are completely determined by the conformal weights of the fields.
Indeed, were it not for the obliteration of the $\beta$'s from the result
the arbitrariness of $\beta$'s noted in \eq{betashift} would have allowed
arbitrary powers of cross ratios in the expression of the conformal 
integrals, hence the correlation functions, rendering the method ineffective. 
Upon choosing appropriate paths of integration the integrals evaluate to
infinite series whose convergence depend on $\xi^A$ \cite{PR2,PR3}.
On the other hand, we can  integrate over the cross ratios over
the real line within the integral.
Let us perform a Mellin transform of $\inmu{0}{N}{\bxi}$ with 
respect to the cross ratios. For $\n=(\nu_1,\nu_2,\ldots,\nu_{N(N-3)/2})$, the
Mellin transform of \eq{I0GG} is 
\begin{equation}
\label{mellin1}
\I_N(\n,\beta,L) = \prod_{A=1}^{N(N-3)/2}\int\limits_0^{\infty} 
	d\xi^A (\xi^A)^{\nu_A-1} \inmu{0}{N}{\bxi}.
\end{equation}
Changing the order of integrations to perform the integrations over the cross
ratios first, followed by integrating over the $t$'s we obtain 
	\begin{equation}
\label{mellin:I}
		\I_N(\n,\beta,L) =(-1)^{\frac{N(N-3)}{2}} \prod_{\substack{i,j=1\\i<j}}^N
		\Ga{\sum\limits_A \nu_A\ell^A_{ij}-\beta_{ij}}.
	\end{equation}

Let us now return to the bootstrap equation.
Plugging the conformal integral \eq{IbI0} in \eq{G4s} and \eq{G4t}, the
bootstrap equation \eq{boot:def} can be written as a product of 
$\inmu{0}{N}{\bxi}$ and powers of the cross ratios. The bootstrap equation,
thereby, is conformal invariant. We can perform Mellin transformation of
all terms of the equation with respect to the cross ratios to arrive
at an equation depending on the OPE coefficients, the $\beta$'s and 
the parameters $\n$. This results in a particularly simple form of the 
bootstrap equation, because, instead of infinite series on both sides of
\eq{boot:def}, only the product of finite number of terms, as in
\eq{mellin:I} appear. Let us point out that the sum over $\D$, remains.

Let us now demonstrate this for the $4$-point bootstrap equation. 
The toric matrix \eq{tormat:def} in the case of $N=4$ is given by \cite{PR2,PR3}
\begin{equation}
\label{amat4}
\mA=\bordermatrix{%
\scriptstyle i\ (jk)&\scriptstyle {(12)} 
&\scriptstyle{(13)}
&\scriptstyle{(14)} 
&\scriptstyle{(23)}
&\scriptstyle{(24)} 
&\scriptstyle{(34)}\cr
\scriptstyle 1 & 1&1&1&0&0&0\cr
\scriptstyle 2 & 1&0&0&1&1&0\cr
\scriptstyle 3 & 0&1&0&1&0&1\cr
\scriptstyle 4 & 0&0&1&0&1&1\cr
}.%
\end{equation}
In order to define the cross ratios \eq{xi:def}, we choose a basis of the
null vectors, arranged into a matrix. In the present instance there are
two null vectors of $\mA$. 
In the $s$-channel the basis null vectors 
are chosen as the transpose of $\ell^1$ and $\ell^2$, given by 
\begin{equation}
        \label{v4}
L= \bordermatrix{%
&\scriptstyle {(12)} 
&\scriptstyle{(13)}
&\scriptstyle{(14)} 
&\scriptstyle{(23)}
&\scriptstyle{(24)} 
&\scriptstyle{(34)}\cr
\ell^1 &0&1&-1&-1&1&0\cr
\ell^2 & 1&0&-1&-1&0&1
}\ %
\begin{matrix}
\xi^1\\\xi^2
\end{matrix}.
\end{equation}
As indicated on the right, the rows of the matrix $L$ define two cross ratios
by taking the entry of each null vector as the exponent of $|x_{ij}|^2$ 
as in \eq{xi:def}, namely,
\begin{equation}
\label{xrat4}
\xi^1=\frac{|x_{13}|^2|x_{24}|^2}{|x_{14}|^2|x_{23}|^2},\quad
\xi^2=\frac{|x_{12}|^2|x_{34}|^2}{|x_{14}|^2|x_{23}|^2}.
\end{equation}
In the $t$-channel, the basis null vectors and cross ratios are similarly 
given by 
\begin{equation}
        \label{v4t}
L'= \bordermatrix{%
&&&&&\cr
\ell'^1 &-1&1&0&0&1&-1\cr
\ell'^2 & -1&0&1&1&0&-1
}\ %
\begin{matrix}
\xi'^1=\xi^1/\xi^2\\\xi'^2=1/\xi^2
\end{matrix},
\end{equation}
which are obtained from $\xi^1$ and $\xi^2$ by exchanging $x_2$ and $x_4$.

The weights $\m$ of the conformal integral appearing in \eq{G4s} and 
\eq{G4t} in the form \eq{IbI0} are used to 
solve for the $\beta$'s, as shown in Table~\ref{tab:4}.
\begin{table}[h]
\setlength\tabcolsep{1em}
\begin{center}
\begin{tabular}{ccc}
\hline
Channel & $\m$ & $\beta$\\
\hline
	$s$-channel~\scalebox{2}{\sChannel} &
$\substack{\mu_1=\frac{\D_1-\D_2+\D}{2}\\
\mu_2=\frac{\D_2-\D_1+\D}{2}\\
\mu_3=\frac{\D_3-\D_4+D-\D}{2}\\
\mu_4=\frac{\D_4-\D_3+D-\D}{2}
}$
&
$\substack{\beta_{12}=0\\
\beta_{14}= \frac{\D_2-\D_1-\D}{2}-\beta_{13}\\
\beta_{23}=\frac{\D_4-\D_3-\D}{2}-\beta_{13}\\
\beta_{24}=\frac{\D_1-\D_2+\D_3-\D_4}{2}+\beta_{13}\\
\beta_{34}=\D-\frac{D}{2}
}$\\ \\
$t$-channel~\scalebox{2}{\tChannel}&
$\substack{
\mu_1=\frac{\D_1-\D_4+\D}{2}\\
\mu_2=\frac{\D_2-\D_3+D-\D}{2}\\
\mu_3=\frac{\D_3-\D_2+D-\D}{2}\\
\mu_4=\frac{\D_4-\D_1+\D}{2}
}$
&
$\substack{
\beta'_{12}=\frac{\D_4-\D_1-\D}{2}-\beta'_{13}\\
\beta'_{14}=0\\
\beta'_{23}=\D-\frac{D}{2}\\
\beta'_{24}=\frac{\D_1-\D_2+\D_3-\D_4}{2}+\beta'_{13}\\
\beta'_{34}=\frac{\D_2-\D_3-\D}{2} -\beta'_{13}
}$
\\
\end{tabular}
\end{center}
\caption{$\beta$'s in the two channels}
\label{tab:4}
\end{table}
OPE consistency \cite{monodrom} of \eq{G4s} and \eq{G4t} fixes
$\beta_{12}=0$ in the $s$-channel and $\beta_{14}=0$ in the $t$-channel. 
Plugging in the values of parameters in \eq{G4s} and \eq{G4t} and equating them
we obtain an equation involving solely the cross ratios, 
\begin{equation}
\label{boot:expl}
(\xi^1)^{\beta_{13}}
\sum_{\D} C_{\D_1\D_2\D}C_{\D\D_3\D_4} (\xi^2)^{\frac{\D-\D_2}{2}}
	\inmu{0}{4}{\xi^1,\xi^2}^{\sChannel}
=
(\xi'^1)^{\beta'_{13}}
\sum_{\D} C_{\D_1\D_4\D}C_{\D\D_3\D_2} (\xi'^2)^{\frac{\D-\D_4}{2}}
	\inmu{0}{4}{{\xi'^1},{\xi'^2}}^{\tChannel},
\end{equation}
as mentioned above.
Let us point out that taking into account the definition of cross ratios
in the two channels, the LHS of the bootstrap equation goes over to the RHS
as $(x_2,\D_2)$ and $(x_4,\D_4)$ are swapped at once.
The GG integral $\inmu{0}{4}{\bxi}$ appears as the summand in the conformal
block. 
Multiplying both sides with $(\xi^1)^{\nu_1-1}(\xi^2)^{\nu_2-1}$,
and integrating over both cross ratios as in \eq{mellin1}
with appropriate variable change on the RHS,
and using \eq{mellin:I}, we obtain 
\begin{multline}
\label{boot4:gen}
\sum_{\D} C_{\D_1\D_2\D}C_{\D\D_3\D_4} 
\I_4\big((\beta_{13}+\nu_1,\tfrac{\D-\D_2}{2}+\nu_2),\beta,L\big) = \\
\sum_{\D} C_{\D_1\D_4\D}C_{\D\D_3\D_2} 
\I_4\big((\beta'_{13}+\nu_1,\tfrac{\D-\D_4}{2}-\nu_1-\nu_2),\beta',L'\big),
\end{multline}
where 
\begin{gather}
\scriptstyle
\I_4\big((\beta_{13}+\nu_1,\tfrac{\D-\D_2}{2}+\nu_2),\beta,L\big) =
\Ga{\nu_1}\Ga{\nu_1-\frac{\D_1-\D_2+\D_3-\D_4}{2}}
\Ga{\frac{\D-\D_2}{2}+\nu_2}
\Ga{\frac{\D_1}{2}-\nu_1-\nu_2}
\Ga{\frac{\D_2+\D_3-\D_4}{2}-\nu_1-\nu_2}
\Ga{\frac{D-\D-\D_2}{2}+\nu_2},
\\
\scriptstyle
\I_4\big((\beta'_{13}+\nu_1,\tfrac{\D-\D_4}{2}-\nu_1-\nu_2),\beta',L'\big)=
\Ga{\nu_1}\Ga{\nu_1-\frac{\D_1-\D_2+\D_3-\D_4}{2}}
\Ga{\frac{\D-\D_4}{2}-\nu_1-\nu_2}
\Ga{\frac{\D_1}{2}+\nu_2}
\Ga{\frac{\D_4+\D_3-\D_2}{2}+\nu_2}
\Ga{\frac{D-\D-\D_4}{2}-\nu_1-\nu_2}.
\end{gather}
Interestingly, the bootstrap equation is independent of the parameters 
$\beta$, only the external weights appear in it. 
This is the general form of the Mellin transform of the 
bootstrap equation depending only on the CFT data.

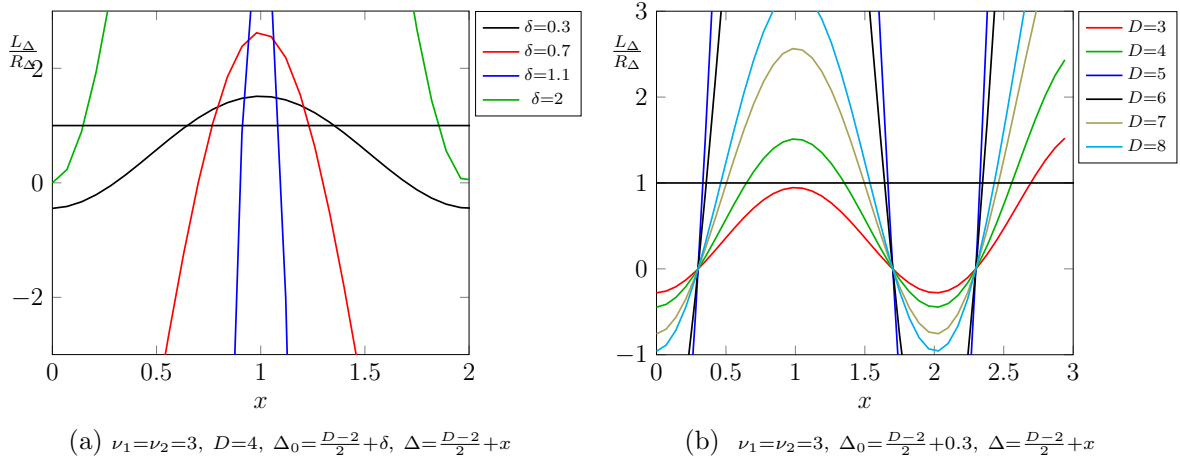
\begin{figure}[h]
\centering
	\subfloat[{$\scriptstyle
\nu_1=\nu_2=3,\ D=4,\ \D_0=\frac{D-2}{2}+\delta,\
\D=\frac{D-2}{2}+x$}
		]{
\label{fig1a}
		\begin{tikzpicture}[scale=.8]
\pgfplotstableread{dzchange.dat}
\dzchange
\begin{axis}[
xmin=0,xmax=2,
ymin=-3,ymax=3,
xlabel=$x$,ylabel=$\frac{L_\D}{R_\D}$,
ylabel style={rotate=-90,at={(-.01,0.88)}},
legend style = { at = {(1.27,1)}}]
\addplot[thick,mark=] table[y = a] from \dzchange ;
\addlegendentry{$\scriptstyle \delta=0.3$}
\addplot[thick,red,mark=] table[y = b] from \dzchange ;
\addlegendentry{$\scriptstyle \delta=0.7$}
\addplot[thick,blue,mark=] table[y = c] from \dzchange ;
\addlegendentry{$\scriptstyle \delta=1.1$}
\addplot[thick,green!70!black,mark=] table[y = d] from \dzchange ;
\addlegendentry{$\scriptstyle \delta=2$}
	\addplot[thick]{1};
\end{axis}
\end{tikzpicture}
	}
	\subfloat[
{$\scriptstyle
\nu_1=\nu_2=3,\
\D_0= \frac{D-2}{2}+0.3,\ 
\D=\frac{D-2}{2}+x
$}]{
\label{fig1b}
		\begin{tikzpicture}[scale=.8]
\pgfplotstableread{dimchange.dat}
\sptimedim
\begin{axis}[
xmin=0,xmax=3,
ymin=-1,ymax=3,
xlabel=$x$,ylabel=$\frac{L_\D}{R_\D}$,
ylabel style={rotate=-90,at={(-.01,0.88)}},
legend style = { at = {(1.25,1)}}]
\addplot[thick,red,mark=] table[y = a] from \sptimedim ;
\addlegendentry{$\scriptstyle D=3$}
\addplot[thick,green!70!black,mark=] table[y = b] from \sptimedim ;
\addlegendentry{$\scriptstyle D=4$}
\addplot[thick,blue,mark=] table[y = c] from \sptimedim ;
\addlegendentry{$\scriptstyle D=5$}
\addplot[thick,mark=] table[y = d] from \sptimedim ;
\addlegendentry{$\scriptstyle D=6$}
\addplot[thick,yellow!60!black,mark=] table[y = e] from \sptimedim ;
\addlegendentry{$\scriptstyle D=7$}
\addplot[thick,cyan,mark=] table[y = f] from \sptimedim ;
\addlegendentry{$\scriptstyle D=8$}
\addplot[thick]{1};
\end{axis}
\end{tikzpicture}
	}
\caption{Bounds on conformal weights}
\label{fig1}
\end{figure}

To demonstrate the usefulness of this form of the bootstrap equation,
let us consider the special case of external fields of equal weight, 
$\D_i=\D_0$, for $i=1,2,3,4$.  
The  bootstrap equation \eq{boot4:gen} simplifies to 
\begin{equation}
\label{boot4:eqwt}
\sum_{\D} C^2_{\D_0\D_0\D} 
(L_\D-R_\D)=0,
\end{equation}
where
\begin{gather}
\label{LR}
L_\D= \Ga{\frac{\D-\D_0}{2}+\nu_2}
\Ga{\frac{D-\D-\D_0}{2}+\nu_2}
\Ga{\frac{\D_0}{2}-\nu_1-\nu_2}^2,\\
R_\D=
\Ga{\frac{\D-\D_0}{2}-\nu_1-\nu_2}
\Ga{\frac{D-\D-\D_0}{2}-\nu_1-\nu_2}
\Ga{\frac{\D_0}{2}+\nu_2}^2.
\end{gather}
This can be used to find bounds on the conformal weights of fields.
Since $C^2_{\D_0\D_0\D}$ is positive, 
a bound on $\D$ is obtained as the ratio $L_\D/R_\D$ crosses unity. 
There are four parameters in the expression, namely, 
$D,\D_0,\nu_1$, and $\nu_2$. The ratio $L_\D/R_\D$ is plotted against
$\D$ in Figure~\ref{fig1}. The weight $\D$ and $\D_0$  are
taken to obey the unitarity bound $\D> (D-2)/2$. In Figure~\ref{fig1a}
the ratio is plotted for different values of $\D_0$, showing that
it crosses unity for some value of $\D$. In Figure~\ref{fig1b} the ratio
is plotted for different dimensions, the graphs crossing unity at a finite
value of $\D$ in each case, thereby providing a lower bound on $\D$. 

%

%

To summarise, we have obtained a simple expression for the Mellin
transform of the $4$-point bootstrap equation for scalar
conformal fields in the $D$-dimensional Euclidean space.  
First, we use the expression of the correlation functions in terms of
conformal integrals in the $s$- and $t$-channel, \eq{G4s} and \eq{G4t},
respectively. Then, writing the conformal integral in the form \eq{IbI0},
the invariant summand of the conformal blocks
depending only on the cross ratios is identified as the
solution of a GG system corresponding to a  toric GKZ system furnished by
the conformal transformations. Writing the bootstrap equation solely 
in terms of cross ratios and using the integral form \eq{mellin1}, we
write down the Mellin transform of the bootstrap equation \eq{boot4:gen}. 
The expression is a sum over the conformal weights of the ``internal" fields, 
$\D$ and has a product of a finite number, $N(N-3)/2$, 
of gamma functions only. 
This formulation avoids dealing with subtleties 
related to the convergence of infinite series,
as well as analytic continuation, in the cross ratios.
Let us note that equation \eq{boot4:gen} is in reality an infinitude 
of equations depending on the Mellin parameters $\n$. 
While we fixed these arbitrarily in Figure~\ref{fig1} to exhibit the 
usefulness of the present formulation, the best bound may depend on particular
values. 

Dealing with a conformal field theory with scalar fields alone is clearly
an over-simplification. However,
the computations presented here can be generalised to incorporate spin 
as well as to bootstrap higher point correlation functions 
\cite{apratim1,apratim2}, since the higher point correlation functions can 
be expressed in terms of conformal integrals, which, in turn, are solutions
of GG systems. Such generalisations will be reported in future. 

\section*{Acknowledgement}
It is a pleasure for me to thank Ant{\'o}nio Antunes,  
Apratim Kaviraj and Kausik Ghosh for patiently sharing their insight.

\end{document}